\documentclass[pre,aps,twocolumn,amsmath,amssymb,showpacs,superscriptaddress]{revtex4}
\usepackage{amssymb}
\usepackage{graphicx}
\usepackage{hyperref}
\usepackage{float}
\usepackage[utf8x]{inputenc}
\usepackage{gensymb}
\usepackage{color}
\begin{document}
\title{Adsorption of interacting self-avoiding trails in two dimensions}

\author{N. T. Rodrigues}\email{nathan.rodrigues@ufv.br}
\affiliation{Departamento de F\'isica, Universidade Federal de Vi\c cosa, 36570-900, Vi\c cosa, MG, Brazil}
\affiliation{School of Mathematical Sciences, Queen Mary University of London, Mile End Road, London, E1 4NS, United Kingdom}
\author{T. Prellberg} \email{t.prellberg@qmul.ac.uk}
\affiliation{School of Mathematical Sciences, Queen Mary University of London, Mile End Road, London, E1 4NS, United Kingdom}
\author{A. L. Owczarek}\email{owczarek@unimelb.edu.au}
\affiliation{School of Mathematics and Statistics, University of Melbourne, Victoria 3010, Australia}

\date{\today}

\begin{abstract}
We investigate the surface adsorption transition of interacting self-avoiding square lattice trails onto a straight boundary line.
The character of this adsorption transition depends on the strength of the bulk interaction,
which induces a collapse transition of the trails from a swollen to a collapsed phase, separated by a critical state.
If the trail is in the critical state, the universality class of the adsorption transition changes; this is known as the special adsorption point.
Using flatPERM, a stochastic growth Monte Carlo algorithm, we simulate the adsorption of self-avoiding interacting trails on the square lattice 
using three different boundary scenarios which differ with respect to the orientation of the boundary and the type of surface interaction.
We confirm the expected phase diagram, showing swollen, collapsed, and adsorbed phases in all three scenarios, and confirm universality 
of the normal adsorption transition at low values of the bulk interaction strength. Intriguingly, we cannot confirm universality of the special 
adsorption transition. We find different values for the exponents; the most likely explanation is that this is due to the presence of strong corrections to scaling at this point.
\end{abstract}

\pacs{05.50.+q,05.10.Ln,05.70.Fh,61.41.+e}

\maketitle

\section{Introduction}
\label{intro}

When a polymer in solution is in contact with an attractive surface, it adsorbes upon decreasing temperature \cite{Eisenriegler1982,DeBell1993,Vrbova1996,Vrbova1998,Vrbova1999,Grassberger2005,Owczarek2007,Luo2008,Klushin2013,Plascak2017}. In the presence of attractive bulk interactions, a polymer collapses from a swollen coil to a collapsed globule \cite{flory1953a-a,gennes1979a-a}. Additionally, a polymer can also undergo a collapse transition when adsorbed on a two-dimensional surface. To complicate matters, introduction of stiffness can give rise to a collapsed crystalline phase \cite{bastolla1997a-a}, giving rise to a complex phase diagram. 

It hence is helpful to consider the simpler scenario of two-dimensional flexible interacting polymers adsorbing onto a line, where we find a two-dimensional phase diagram with three phases: swollen coil, collapsed globule, and adsorbed polymer \cite{veal1991a-a}. The canonical lattice model for this is given by self-avoiding walks on the square lattice tethered to a point on the surface, with energetic contributions from the number of non-consecutive nearest-neighbor sites of the walk (bulk interactions) and the number of sites of the walk in the surface (surface interactions). This model has been extensively studied previously, and theoretically predicted critical exponents have been confirmed numerically with high accuracy \cite{foster1992a-a}.

Recently, there has been renewed interest in the polymer adsorption transition. In \cite{Plascak2017,Martins2018}, it was argued based on numerical simulations in three dimensions, that the generally accepted scaling theory for polymer adsorption in terms of a single crossover exponent may break down in the presence of bulk interactions. Specifically, it was claimed that the exponent involved in the temperature scaling of the free energy around the adsorption critical point is distinct from the exponent describing the scaling of the order parameter at this critical point, and moreover that these two exponents are not universal with respect to varying the strength of the bulk interactions.

Thus, there was need to examine the adsorption transition more carefully in a variety of two- and three-dimensional models \cite{Bradly2018, Bradly2018a}, in order to both test the universality assumption as well as the validity of the numerical methods used to estimate the scaling exponents. The main conclusion of that work was that existing methods for extracting critical exponents for the adsorption transition from numerical data do not seem to be able to capture the effect of finite-size corrections. Different methods produced different exponent estimates with statistical errors much smaller than the difference between the estimates, and there also was no clear indication as to which method was most reliable. It therefore seemed likely that any apparent non-universal behaviour was due to the fact that the methods used could not account sufficiently well for systematic error.

One of the models studied in \cite{Bradly2018} was the model of self-avoiding trails on the square and simple cubic lattice, weighted by the number of sites in the surface. Trails are lattice paths which may repeatedly visit sites but traverse bonds only once. If one associates a contact interaction to every multiply-visited lattice site, then one can view self-avoiding walks as self-avoiding trails with infinite repulsion. It is known that self-avoiding trails and self-avoiding walks on the square lattice are in the same universality class, with subtle differences in the corrections to scaling \cite{guim1997a-a}. Interacting self-avoiding trails on the square lattice undergo a collapse transition, which however is not in the same universality class as the collapse transition of interacting self-avoiding walks \cite{Owczarek1995,foster2009a-a}. 

In the current paper we extend previous studies by considering the adsorption transition for interacting self-avoiding trails on the square lattice. We analyse the adsorption transition in the presence of bulk interactions of varying strength. In addition to the normal adsorption transition from the swollen to the adsorbed phase, we also study the special surface transition occurring when collapsing polymers adsorb. The value of the bulk interaction at which interacting self-avoiding trails on the square lattice collapse is exactly known, as is the free energy at this point \cite{Owczarek1995}, which is a major advantage of studying this model in contrast to interacting self-avoiding walks, where only a numerical estimate of the collapse transition point is available.

The temperature of the adsorption transition is sensitive to the orientation of the boundary and type of boundary interaction, but the same universal critical exponents are expected. However, previous works \cite{Owczarek1995,Foster2010} which have considered different surface interactions are in disagreement about the value of the exponents for the special transition. Alternatively to considering the number of sites of a trail in the surface \cite{Bradly2018}, one can consider the number of bonds of a trail in the surface \cite{Foster2010}. Also, while conventionally a horizontal surface is considered, for interacting self-avoiding trails at the collapse point it is advantageous to consider weighting the number of sites along a diagonal line, as in this case the value of the surface interaction at which collapsing trails adsorb is exactly known \cite{Owczarek1995}. We hence investigate normal and special adsorption for all three scenarios: a horizontal boundary with either site or bond interactions, and a diagonal boundary with site interactions.

The paper is structured as follows. In Section \ref{defmod} we define the lattice model we investigate, and introduce relevant thermodynamic quantities. In Section \ref{scaling} we review scaling laws and critical exponents, and describe the methods by which we extract exponents from numerical data. In Section \ref{Simul} we describe the simulation methods used in this work, and Section \ref{results} describes our findings in detail. 

\onecolumngrid

\begin{figure}[!h]
\includegraphics[angle=0, width=.8\columnwidth]{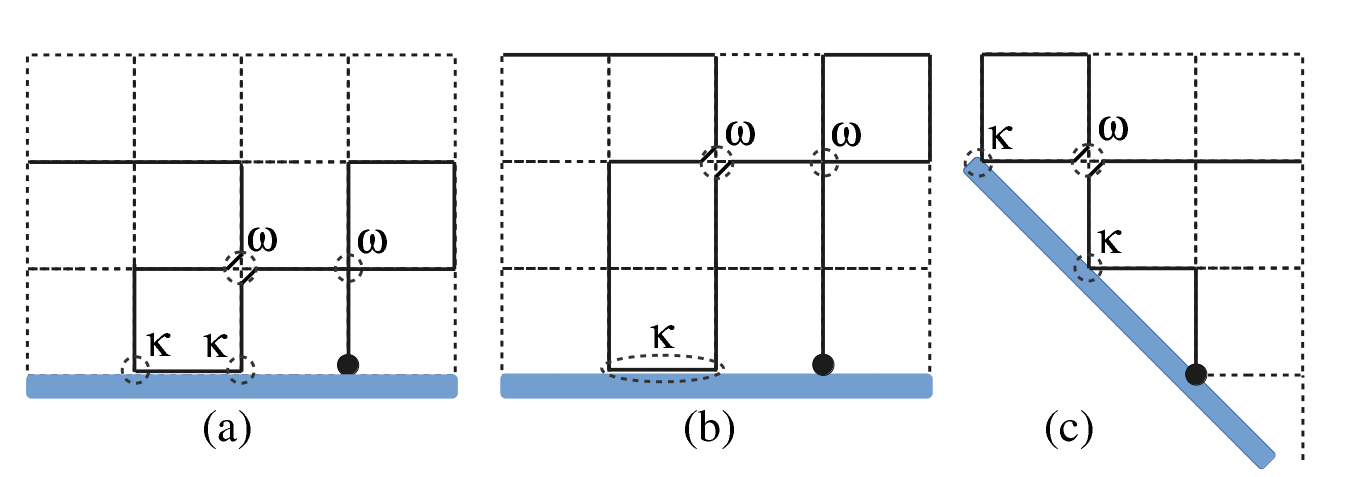}
\caption{Examples of trails for the three cases of surface interaction studied. The trail starts at the solid circle in the surface. Bulk and surface interactions are indicated by dashed circles and are denoted by $\omega$ and $\kappa$, respectively. Panel (a) shows monomer-surface interactions in a horizontal surface, panel (b) shows bond-surface interactions in a horizontal surface, and panel (c) shows monomer-surface interactions in a diagonal surface. Note that the starting point does not contribute for the surface energy.}
\label{Fig1}
\end{figure}
\twocolumngrid

\section{The Model}
\label{defmod}

A self-avoiding trail (SAT) is a finite lattice path on a regular lattice in which every bond may only be traversed once. We identify the monomers of a polymer with the site visits of the path, and allow for more than one monomer on a site. In this work we are interested in the adsorption and collapse transition in the SAT model on the square lattice. To achieve this we need to introduce two types of interaction: a bulk interaction $\epsilon_b$ between doubly visited sites (i.e. monomers on the same site), and a surface interaction $\epsilon_s$ for a monomer (or bond) lying in the surface, which we take to be the boundary of a half-plane. We fix the boundary to contain the origin and consider trails starting at the origin, i.e.~on the boundary.

We consider two ways to define the interaction between a trail and the surface, given by either the monomers or bonds of the lattice path in the surface. We shall denote this the monomer-surface or bond-surface case. Examples of this are shown in panels (a) and (b) of Fig.~\ref{Fig1}, respectively, where the surface is aligned with one of the lattice directions. When the surface is not aligned, then bonds cannot lie in the surface, and it only makes sense to consider the monomer-surface case. An example of a surface oriented at $45\degree$ is shown in panel (c) of Fig.~\ref{Fig1}.

The canonical (fixed length) partition function for adsorbing and interacting trails is given by
\begin{equation}
Z_n(\kappa,\omega) = \sum_{m_s,m_b}C^{(n)}_{m_s,m_b} \kappa^{m_s} \omega^{m_b},
\end{equation}
where $\kappa=e^{\beta\epsilon_s}$, $\omega=e^{\beta\epsilon_b}$ and $C^{(n)}_{m_s,m_b}$ is the number of $n$-step lattice paths with $m_s$ surface contacts and $m_b$ doubly visited sites, and $\beta=1/{k_bT}$ with $T$ the temperature. The reduced finite-size free energy is
\begin{equation}
	f_n(\kappa,\omega)  = - \frac{1}{n} \log Z_n(\kappa,\omega)\;,
\end{equation}
which in the thermodynamic limit gives
\begin{equation}
	f_\infty(\kappa,\omega) = \lim_{n\rightarrow\infty} f_n(\kappa,\omega)\;.
\end{equation}
Any general thermodynamic quantity $Q$ gives rise to averages
\begin{equation}
    \langle Q \rangle(\kappa,\omega) = \frac{1}{Z_n(\kappa,\omega)}\sum_{\psi_n} \kappa^{m_s(\psi_n)}\omega^{m_b(\psi_n)} Q(\psi_n)\;,
    \label{eq:ThermoQuantity}
\end{equation}
where the sum ranges over all $n$-step lattice trails $\psi_n$.
In particular, we are interested in the surface internal energy
\begin{equation}
    u_n (\kappa,\omega) = \frac{\langle m_s \rangle}{n}\;.
    \label{eq:InternalEnergy}
\end{equation}
As this can be interpreted as the fraction of the trail that is adsorbed in the surface, it is an order parameter for the surface adsorption transition.

We also consider the components of the mean-squared end-to-end radius $ R^2_n$ parallel and perpendicular to the surface. For a horizontal surface these are defined as
\begin{align}
    R^2_{\perp,n} (\kappa,\omega)&= \langle x_n^2 \rangle, \\
    R^2_{\parallel,n} (\kappa,\omega)&= \langle y_n^2 \rangle,
    \label{eq:EndToEndRadius}
\end{align}
with the endpoint of the trail being at $(x_n,y_n)$, whereas for a diagonal surface they are defined as
\begin{align}
    R^2_{\perp,n} (\kappa,\omega)&= \frac12\langle(x_n+y_n)^2\rangle, \\
    R^2_{\parallel,n} (\kappa,\omega)&= \frac12\langle(x_n-y_n)^2\rangle.
    \label{eq:EndToEndRadius_diag}
\end{align}

\section{Scaling Laws and Critical Exponents}
\label{scaling}
The surface internal energy $u_n$ is the order parameter of the adsorption transition. For long lengths at the critical point $u_n$ scales as $u_n\sim n^{\phi^{(a)}-1}$. For finite lengths, finite-size corrections need to be included:
\begin{equation}
u_n \sim n^{\phi^{(a)}-1}f^{(0)}_u(X)\left[1 + n^{-\Delta}f_u^{(1)}(X) + ...\right],
\label{un}
\end{equation}
where $f_u^{(i)}$ are scaling functions of the variable $X=(T_a-T)n^{1/\delta}$ and $\Delta$ is the first correction term.

As a consequence, for a fixed value of $X$ (i.e.~near a critical transition), this induces a relationship between $T$ and $n$ and therefore we can infer a dependence of the finite-size transition temperature $T_a^{(n)}$ of the form 
\begin{equation}
T_a^{(n)} \sim T_a + \text{const}\phantom{.}n^{-1/\delta},
\label{Ta}
\end{equation}
and hence $1/\delta$ is identified as the crossover exponent for the adsorption transition. Both critical exponents, ${\phi^{(a)}}$ and $1/\delta$ are believed to be universal and equal to each other~\cite{Bradly2018}. For the normal surface transition in two dimensions it is expected that $\phi^{(a)}=1/\delta=1/2$.

In what follows, we need to modify our notion of temperature to take into account \emph{only} the surface interactions, and not the bulk interactions. We accomplish this by formally introducing two temperature variables by writing $\kappa=e^{\beta\epsilon_s}=e^{1/T_s}$ and $\omega=e^{\beta\epsilon_b}=e^{1/T_b}$. For the adsorption transition, we fix $T_b$ and in a slight abuse of notation identify $T$ with $T_s$. We can then measure $1/\delta$ independently of $\phi^{(a)}$ by calculating the logarithmic derivative of $u_n$:
\begin{equation}
\Gamma_n 
= \frac{d \log u_n}{dT_s} 
= (\log\kappa)^2 \frac{\left<m_s^2\right> - \left<m_s\right>^2}{\left<m_s\right>}\;.
\end{equation}
$\Gamma_n$ is related to a second derivative of the free energy, therefore the peaks of $\Gamma_n$ have the following scaling form:

\begin{equation}
\max \Gamma_n \sim n^{1/\delta}f^{(0)}_\Gamma(X)\left[1 + n^{-\Delta}f_\Gamma^{(1)}(X) + ...\right]\;.
\label{Gamma}
\end{equation}
Using the dependence of $\max \Gamma_n$ we can therefore determinate the adsorption transition temperature and the exponent $1/\delta$.

Another way to determine the adsorption point is using metric quantities. Using the scaling behaviour of the 
 parallel and the perpendicular components $R^2_{\perp/\parallel,n}$ with respect to the surface
\begin{equation}
\label{r2}
R^2_{\perp/\parallel,n}\sim n^{2\nu_{\perp/\parallel}},
\end{equation}
where $\nu_{\perp/\parallel}$ is the respective Flory exponent, we can calculate finite-size estimates of these exponents simply by using Eqn. (\ref{r2}):
\begin{equation}
\label{nu}
\nu_{\perp/\parallel,n}=\frac{1}{2\log 2}\log\left(\frac{R^2_{\perp/\parallel,n}}{R^2_{\perp/\parallel,n/2}}\right)\;.
\end{equation}
In the desorbed phase both components have the same value in the thermodynamic limit. For an adsorbed configuration the polymer becomes a quasi-one-dimensional system and $\nu_\perp \rightarrow 0$ while $\nu_\parallel\rightarrow 1$. For some intermediate temperature the components of $\nu$ cross, and using these crossing points we can locate the finite-size temperatures of adsorption $T_a^{(n)}$.

Similarly, we can use the asymptotic scaling of $R^2_n$ to determine and locate the collapse transition point as $\omega$ changes in the desorbed regime, i.e.~ for small values of $\kappa$. At high temperatures $\nu$ assumes the Flory value of $\nu=3/4$ for the swollen phase, and for low temperatures the value of $\nu=1/2$ for the collapsed phase. At the collapse point a transition occurs and the exponent $\nu$ assumes a different value. In the literature one can find $\nu=12/23$ \cite{foster2009a-a} and also $\nu=1/2$ with the presence of logarithmic corrections \cite{Owczarek1995} (this is different from interacting self-avoiding walks, where the value $\nu_{\theta}=4/7$ is well established \cite{Duplantier1987}). We can estimate the finite-size collapse temperature and the corresponding exponent $\nu$ by locating the crossing point in $\nu_n(\omega)$ curves for different lengths. Note that while for the adsorption transition we considered the crossing of exponent estimates from two different components at the same length, here we consider the crossing of exponent estimates of different lengths.

While for the adsorption transition we use $\Gamma_n$ to find the crossover exponent, in the collapse transition the quantity of interest is the bulk specific heat per monomer: 
\begin{equation}
c_n(T_b) = \frac{(\log w)^2}{n}\left(\left<m_b^2\right> - \left<m_b\right>^2\right)\;.
\label{cn}
\end{equation}

Around the collapse temperature a tricritical crossover scaling form is expected. Assuming that in the thermodynamic limit the specific heat diverges as $c(T_b)\sim|T_b-T^{(c)}|^{-\alpha}$, and assuming that the tricritical scaling relation
\begin{equation}
2 - \alpha = \frac{1}{\phi^{(c)}}
\end{equation}
holds, the finite size specific heat $c_n(T_b)$ has the following scaling:
\begin{equation}
c_n(T_b)\sim n^{2\phi^{(c)}-1}f^{(0)}_c(Y)\left[1 + n^{-\Delta}f_c^{(1)}(Y) + ...\right]\;,
\label{cnScl}
\end{equation}
where $Y=(T_b - T^{(c)})n^{\phi^{(c)}}$. The exponent $\phi^{(c)}$ is the crossover exponent for the collapse transition.

\section{Numerical simulations}
\label{Simul}

We sample trail configurations using the flatPERM algorithm \cite{Prellberg2004}. This method is an extension of the Pruned-Enriched Rosenbluth method (PERM) \cite{Grassberger1997}. Both PERM and flatPERM are stochastic growth algorithms based on the Rosenbluth method; the addition of pruning and enrichment allows to overcome attrition and to efficiently sample large configurations. PERM gives an estimate of the partition function $Z_n$ for a specific temperature while flatPERM samples a flat histogram giving a good estimate of the density of states, i.e., the number of configurations $C^{(n)}_{m_s,m_b}$.

We perform simulations for three different scenarios, (a) a horizontal boundary with monomer-surface (MS) interactions, (b) a horizontal boundary with bond-surface (BS) interactions, and (c) a diagonal surface (DS) with monomer-surface interactions. In all three scenarios we first run a two-parameter flatPERM simulation. In this case the algorithm samples a flat histogram in both $m_s$ and $m_b$ (as well as $n$ up to a maximal length) and estimates the full density of states at these lengths, allowing us to construct a finite-size approximation to the phase diagram. 

The two-parameter flatPERM simulation produces a two-dimensional density of states, and it is necessary  to generate sufficiently many samples for each box of the histogram. Therefore this is
only feasible for relatively short lengths. For all three scenarios we perform a two-parameter flatPERM simulation for trails with up to $128$ steps with $10^{10}$ trails reaching the maximum length. From these simulations we can then determine the different phases and the approximate location of phase boundaries.

To determine more precisely the location of and the behaviour around the phase boundaries we perform one-parameter flatPERM simulations for fixed values of $\omega$ or $\kappa$ and generating a one-dimensional density of states for $m_s$ or $m_b$, respectively. This allows for a more detailed analysis in specific regions of the phase diagram. As we only need to generate a one-dimensional density of states, we can perform simulations for longer lengths than for the two-parameter flatPERM simulations. We can generate trails with up to $1024$ steps with $10^{10}$ trails reaching the maximum length for a wide range of $\omega$ and $\kappa$.

As we want to pay particular attention to the normal and the special surface transitions, we also perform PERM (i.e. zero-parameter flatPERM) simulations for all three scenarios, as this allows us to perform simulations for much larger lengths. We generate trails with length up to $10240$ steps with an average sample of $5\times10^8$ trails at maximum length for a set of fixed values of $\omega$ and $\kappa$. At these large lengths the finite-size corrections to scaling are significantly smaller than at shorter lengths, which allows for a more reliable estimate of the adsorption exponents for both ordinary and special surface transitions.

\section{Results}
\label{results} 
For all three scenarios we first generate a finite-size approximation to the phase diagram. Transition regions between different phases are characterised by large fluctuations in the number of bulk and surface interactions $m_b$ and $m_s$. To identify the regions of maximal fluctuations, it is advantageous to consider the covariance matrix 
\begin{equation}
\begin{bmatrix} 
\left<m_s^2\right> - \left<m_s\right>^2 & \left<m_s m_b\right> - \left<m_s\right>\left<m_b\right> \\
\left<m_s m_b\right> - \left<m_s\right>\left<m_b\right> & \left<m_b^2\right> - \left<m_b\right>^2 
\end{bmatrix}
\label{Mx}
\end{equation}
and calculate its largest eigenvalue. We produce finite-size fluctuation maps by plotting the logarithm of this eigenvalue as a function of $\omega$ and $\kappa$ for trails with $n=128$ steps.

\begin{figure}[!h]
\includegraphics[angle=0, width=0.8\columnwidth]{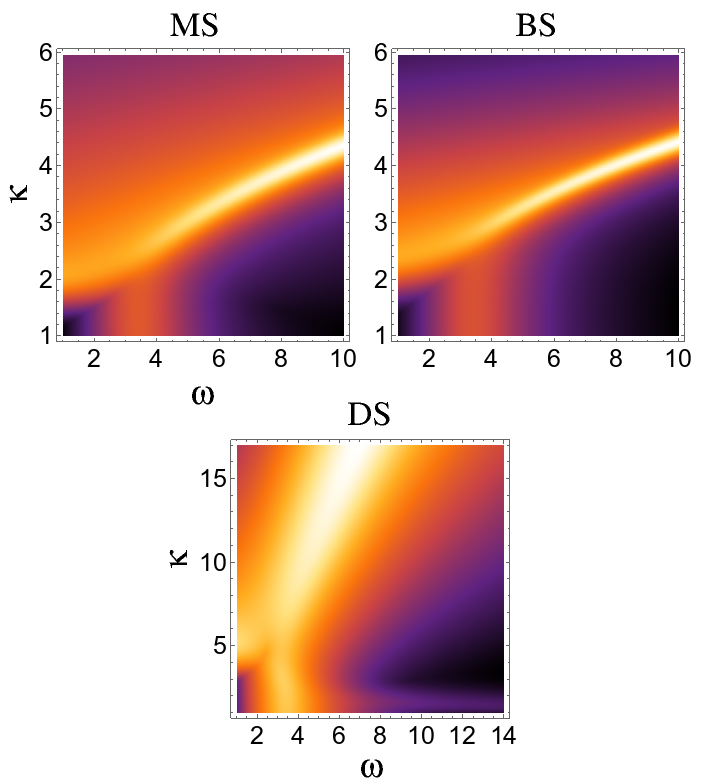}
\caption{Finite size fluctuation map for $128$-step trails for the three boundary cases MS, BS, and DS. Darker colors represent regions of small fluctuations, while brighter colors (yellow/orange) represent regions with strong fluctuations.}
\label{Fig2}
\end{figure}

In Fig.~\ref{Fig2} the logarithm of the largest eigenvalue is shown in a density plot as a function of $\omega$ and $\kappa$. As expected, in each scenario we find three phases, which by considering averages of $m_s$ and $m_b$ we identify with the coil, collapsed and adsorbed states of the trail.

Qualitatively, the phase diagrams in the three scenarios are similar, and we therefore only discuss the MS case in detail, which is shown in Fig.~\ref{Fig2} (a). For small values of $\omega$ and $\kappa$ we find configurations dominated by a small number $m_s$ of surface contacts and a small number $m_b$ of double visited sites. We therefore conclude that this region can be identified with the swollen coil phase. 

When increasing $\omega$ while keeping $\kappa$ constant at a small value ($\kappa\lesssim 2$) the number of double visited sites increases through the region $2<\omega<5$ and gets saturated for large $\omega$, where the trail configurations are dominated by a large number $m_b$ of double visited sites, which is a characteristic of the collapsed phase in this model. We therefore conclude that there is a collapse transition, which for short trail lengths is smoothed out over a wide range. 

\onecolumngrid

\twocolumngrid

For small values of $\kappa$, the average number $m_s$ of surface contacts is small, and the trail remains desorbed, but for large values of $\kappa$ we find an adsorbed phase characterised by trail configurations with a large number $m_s$ of surface contacts. This phase exists for all values of $\omega$, and we therefore have a transition between the desorbed and adsorbed regime. The transition between the desorbed phases and the adsorbed one can occur in three different ways.

For small values of $\omega$ there is a weak transition between the swollen coil and the adsorbed phase which is known as the ``normal surface transition'', and for large values of $\omega$ there is a strong transition between the collapsed and the adsorbed phase. For the latter transition we find a bimodal density of states which is indicative of a first-order transition. Between these two different transitions we expect to see another adsorption transition when increasing $\kappa$ along the line of critical collapse. This transition is known as the ``special surface transition''. It occurs at a multi-critical point at which three transition lines meet: the normal surface, the coil-collapsed and the collapsed-adsorbed lines.

\begin{figure}[!h]
\includegraphics[angle=0, width=0.7\columnwidth]{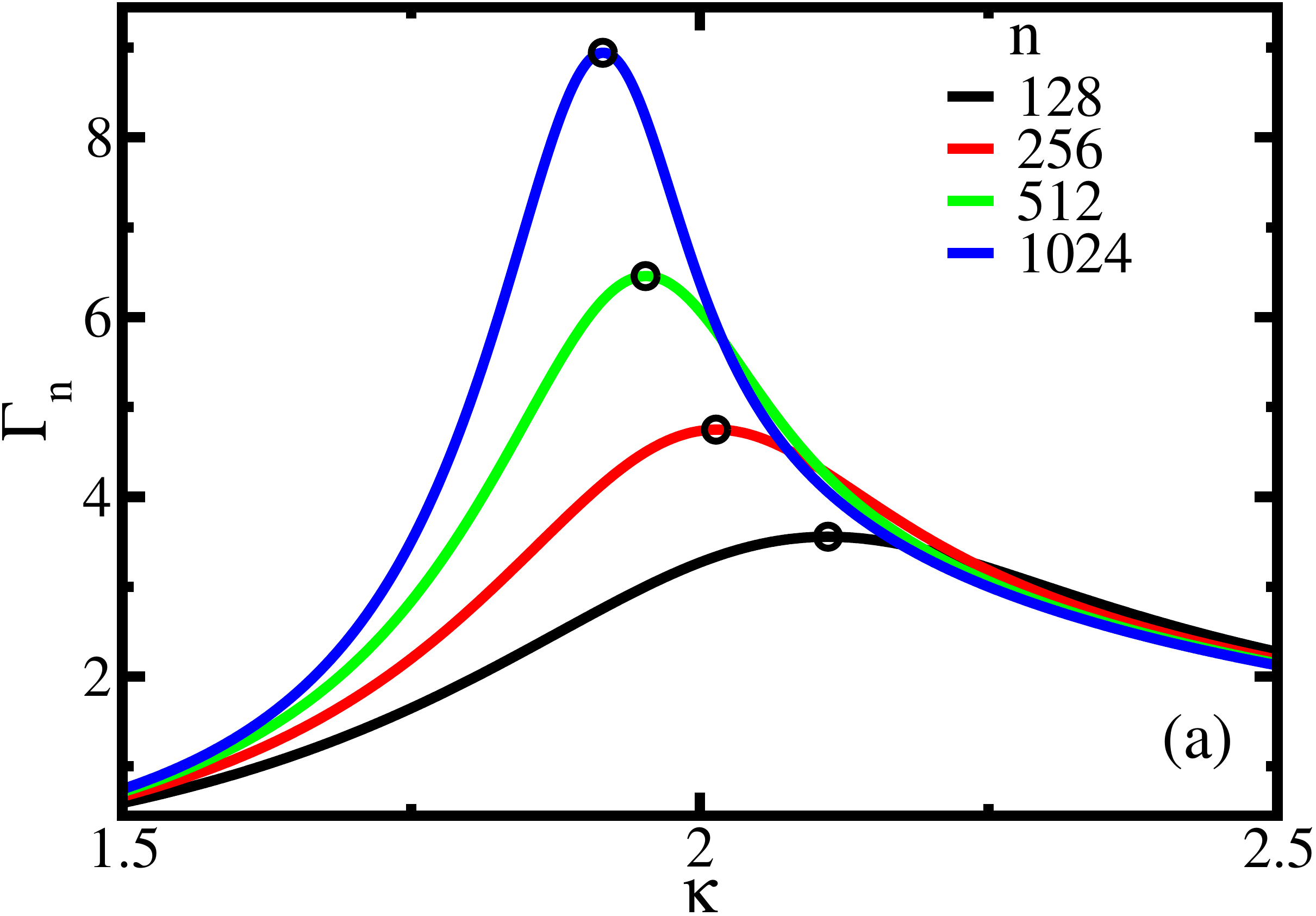}
\includegraphics[angle=0, width=0.7\columnwidth]{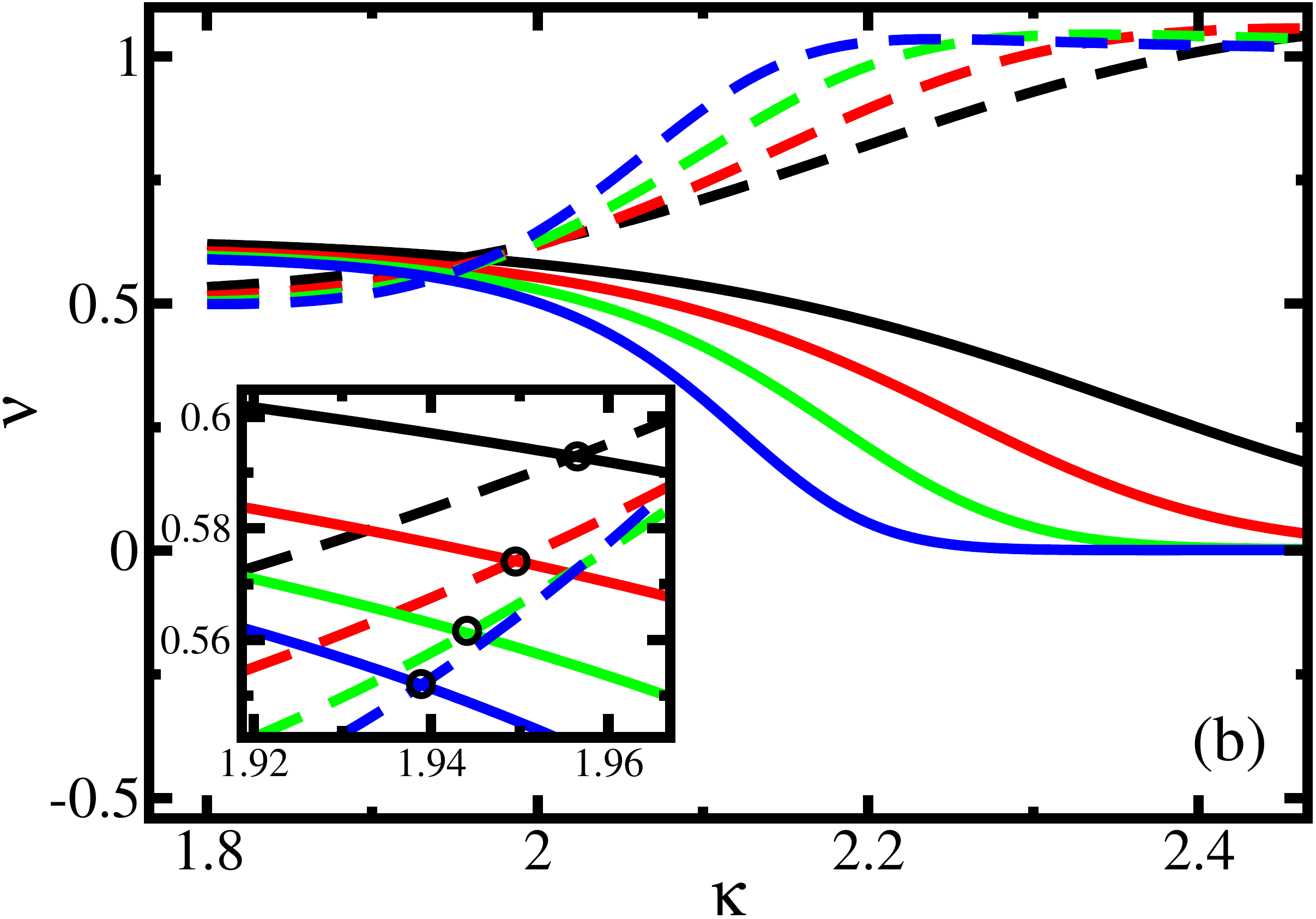}
\includegraphics[angle=0, width=0.7\columnwidth]{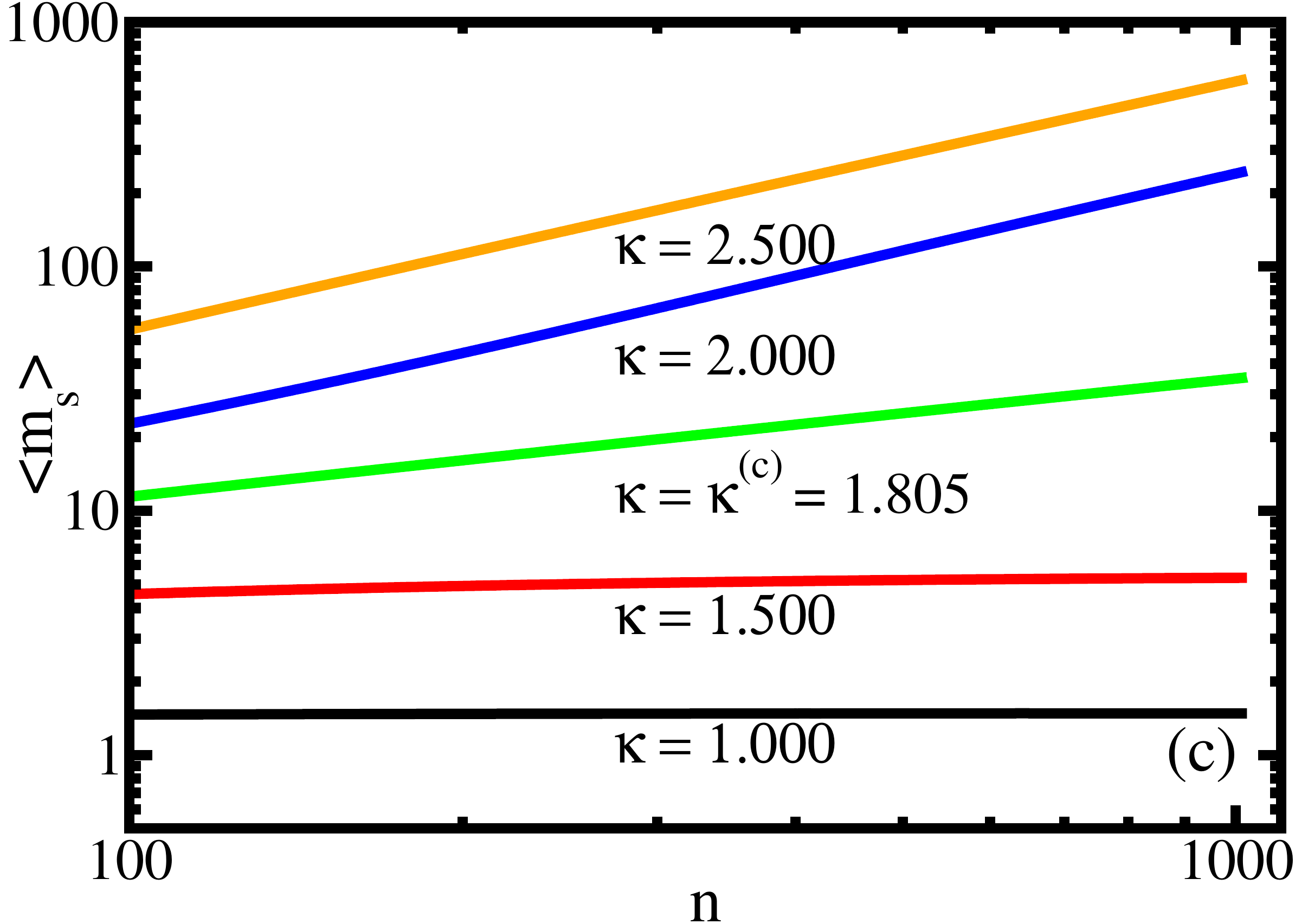}
\includegraphics[angle=0, width=0.7\columnwidth]{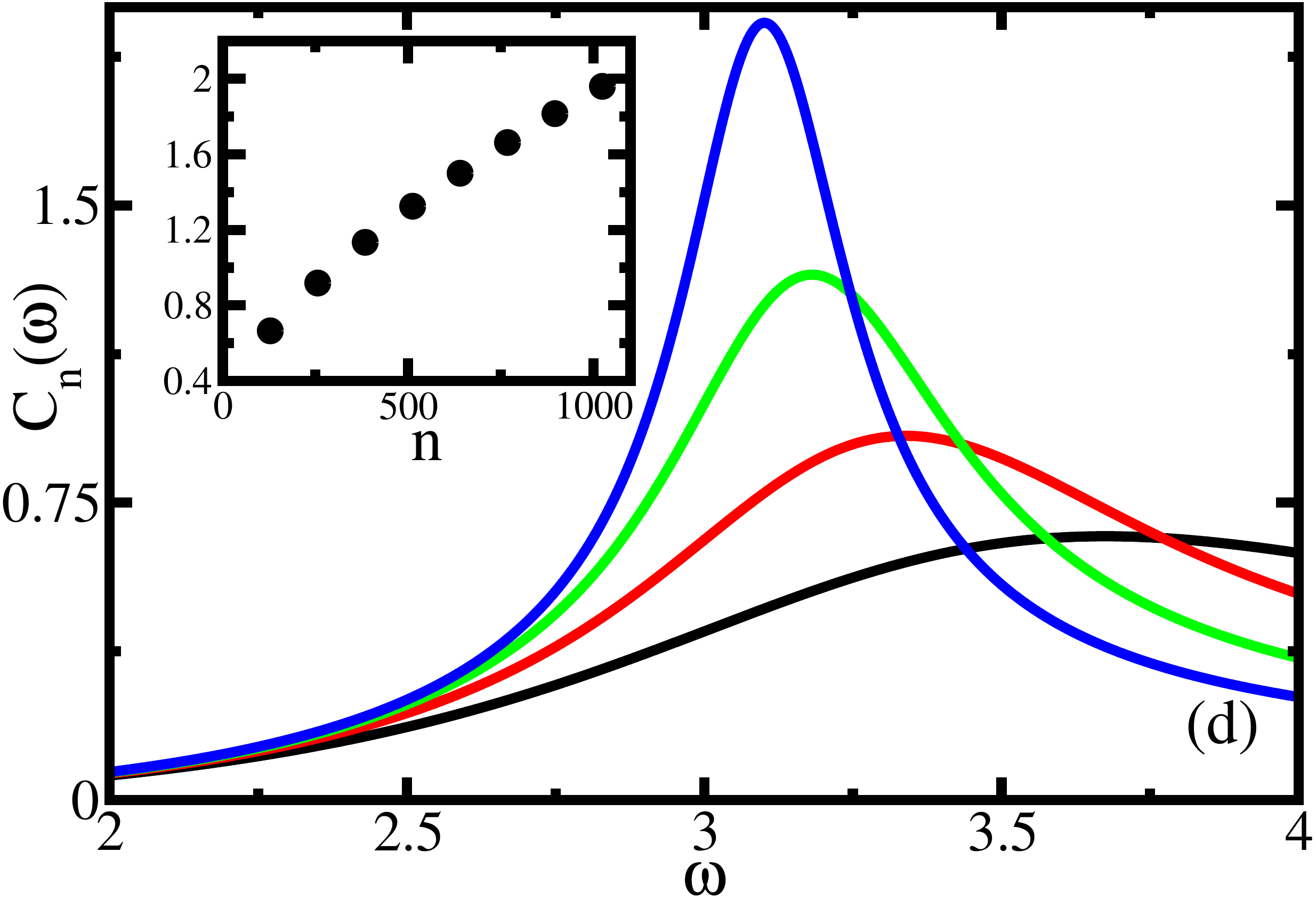}
\caption{(a) Graphs of $\Gamma_n$ as a function of $\kappa$ for $\omega=2$. The black circles denote the maximum points $(\kappa_n,\max\Gamma_n)$ of the $\Gamma_n$ curves. (b) Graphs of the length scale exponents $\nu_{\perp,n}$ (solid line) and $\nu_{\parallel,n}$ (dashed line) as a function of $\kappa$ for $w=3$. The inset shows the crossing points (black circles) of these exponents. (c) The average number of contacts $\left<m_s\right>$ as function of $n$ for $\omega=1$ for five different values of $\kappa$. (d) The bulk specific heat per monomer for $\kappa=1$. The inset shows the location of the maximum values of the bulk specific heat as function of the length $n$.}
\label{Fig3}
\end{figure}

In order to estimate the location of the transitions and the associated critical exponents we simulate configurations of longer lengths by fixing  either $\omega$ or $\kappa$., and use the following procedure for the analysis of the data. For the adsorption transition we choose a fixed value of $\omega$ and find the maximum point $(\kappa_n,\max\Gamma_n)$ as a function of $n$, shown as small circles in Fig.~\ref{Fig3} (a). We then use the asymptotic behaviour of Eq.~\ref{Gamma} to find the value of the exponent $1/\delta_n$. By extrapolating these values to large $n$ we then estimate the final value of $1/\delta$. As shown in \cite{Bradly2018,Bradly2018a}, a good way to estimate the adsorption point is to find the crossing points of the parallel and the perpendicular length scale exponents $\nu_{\perp/\parallel,n}$ (Fig.~\ref{Fig3} (b)). Finding these points and using the scaling of Eq.~\ref{Ta} and the value of $1/\delta$ previously estimated, we locate the transition point ($\omega^{(a)},\kappa^{(a)}$). Knowing this adsorption point, we then calculate the average number of surface contacts $\left<m_s\right>$ (Fig.~\ref{Fig3} (c)) at ($\omega^{(a)},\kappa^{(a)}$) as a function of $n$. As expected, $\left<m_s\right>$ grows linearly in the adsorbed regime and tends to a constant in the desorbed regime, while following a distinct power law growth with an exponent around $0.5$ at the adsorption transition. Similarly our method of estimating $1/\delta$ we use Eqs.~\ref{eq:InternalEnergy} and \ref{un} to find the finite-size values of $\phi^{(a)}_n$ and then we extrapolate these values to estimate the value of $\phi^{(a)}$.

To characterize the collapse transition for a fixed value of $\kappa$ we first calculate the bulk specific heat per monomer (Eq.~\ref{cn}) as a function of $\omega$. By looking at the maxima (Fig.~\ref{Fig3} (d)) and using the equation Eq.~\ref{cnScl} we estimate the exponent $\phi^{(c)}$. To locate the collapse transition point we proceed similar to the adsorption transition, but we consider crossing points ($\omega^{(c)}_n$, $\nu^{(c)}_n$) of the graphs of $\nu_n$ as a function of $\omega$ for two lengths $n$ and $n+\Delta n$. Using the value of $\phi^{(c)}$ we estimate 
$\omega^{(c)}$ by extrapolating from finite-size estimates $\omega^{(c)}_n$. 

In the asymptotic estimation of the exponents above we assume that the finite size estimates $\eta_n$ of an exponent $\eta$ asymptotically satisfies the Ansatz
\begin{equation}
\eta_n = \eta_{\infty} + \text{const } n^{-0.5} + \ldots\;.
\end{equation}
In most cases, this Ansatz appears to fit our data reasonably well, and changing the power in the correction-to-scaling term slightly does not seem to affect our results.

\begin{figure}[!h]
\includegraphics[angle=0, width=0.7\columnwidth]{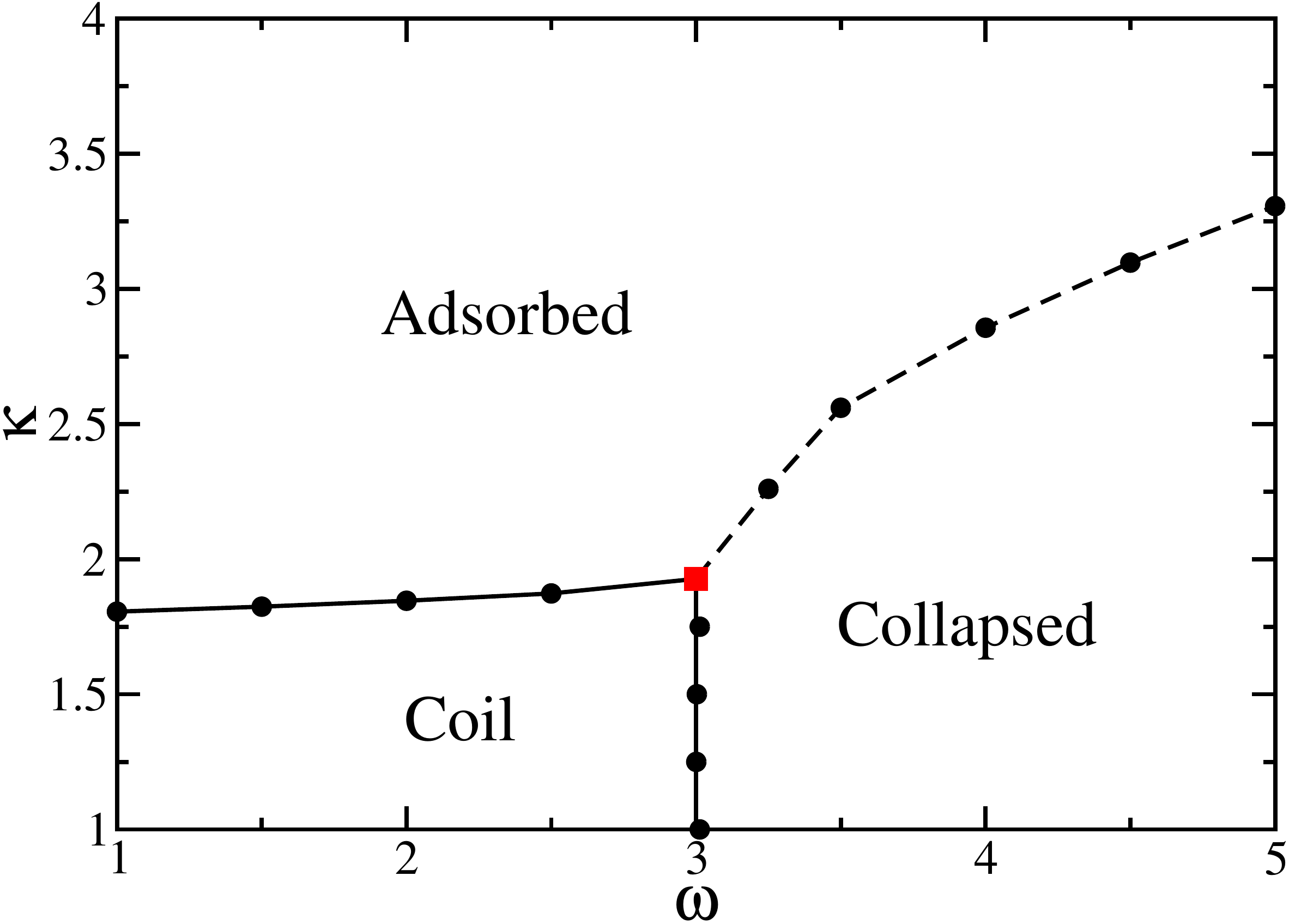}
\caption{The phase diagram for the MS case. The red square is the location of a multicritical point where the special surface transition occurs. The solid lines are critical transitions, i.e. the coil-collapsed transition and the normal surface transition. The dashed line is the coexistence line of the collapsed-adsorbed transition.}
\label{Fig4}
\end{figure}

For the MS case we applied this procedure to the results of the one-parameter flatPERM simulation for trails with up to $1024$ steps in the range of $1\leqslant \kappa \lesssim 3.5$ and $1\leqslant \omega \leqslant 5$. In Fig.~\ref{Fig4} the phase diagram for the MS case is shown. 

On the square lattice, the collapse transition for trails is expected to occur at $\omega^{(c)}=3$~\cite{Owczarek1995}. At this value, the probabilities of a stochastic growth process are perfectly balanced by the Boltzmann weight $\omega^{(c)}$. This has also been confirmed numerically \cite{Foster2010,foster2009a-a}, where a value of $3.000(1)$ was found.
The presence of a weakly interacting surface is not expected have any effect on the location of this transition. Upon increasing the strength of the surface interaction $\kappa$, the collapse transition remains at $\omega^{(c)}=3$ until a multi-critical point is reached at $\kappa=\kappa^{(s)}$, where the special surface transition takes place. 
Our simulations of interacting trails in the presence of a non-interacting surface (i.e. $\kappa=1$) gives a value of $\omega^{(c)}=3.013(10)$ for the MS case, and values very close to $3$ were also found for the BS and the DS cases. We further confirm that the location of the collapse transition does not change upon increasing $\kappa$, as shown in
Fig.~\ref{Fig4}.

Together with estimating the location of the collapse transition, we also obtain estimates of the length-scale exponent $\nu^{(c)}$ at collapse, as well as the collapse crossover exponent $\phi^{(c)}$. On the line $\omega=\omega^{(c)}$ at the values of $\kappa$ indicated in Fig.~\ref{Fig4}., we find $0.538< \nu^{(c)} <0.560$ and $\phi^{(c)}$ very close to $0.78$. The values found for $\nu$ match well with finite-size estimates from the data presented in ~\cite{Owczarek1995} at corresponding lengths when assuming simple power law scaling. They are not close to $\nu^{(c)}=12/23$ reported in \cite{foster2009a-a} or $\nu^{(c)}=1/2$ found in~\cite{Owczarek1995}, but rather indicative of strong finite-size corrections to scaling at the collapse transition. The value of the collapse crossover exponent $\phi^{(c)}$ is also not close to the expected $\phi^{(c)}\approx 0.88$~\cite{Owczarek1995}, but mirror what was found for similar lengths in \cite{meirovitch1989b-a}, again indicative of strong finite-size corrections.

We now turn to the discussion of the adsorption transition in the three regimes. For fixed $\omega>3$, upon increasing $\kappa$ we find a collapsed-adsorbed transition with a bimodal behaviour in the density of states and an exponent $\alpha$ close to $1$, which is a clear indication of a first order transition. For $\omega\leqslant 3$ we observe a critical adsorption transition upon increasing $\kappa$. For the normal surface transition ($\omega<3$) the surface exponents $1/\delta$ and $\phi^{(a)}$ are expected to be equivalent and equal to $1/2$ in two dimensions. For the special surface transition at $\omega=3$, however, a different value is expected. For the DS case, the value $\phi^{(s)}\approx 0.44$ was found previously in~\cite{Owczarek1995}, and for the BS case values slightly lower were reported: $0.379<\phi^{(s)}<0.414$~\cite{Foster2010}.

\begin{figure}[!h]
\includegraphics[angle=0, width=0.7\columnwidth]{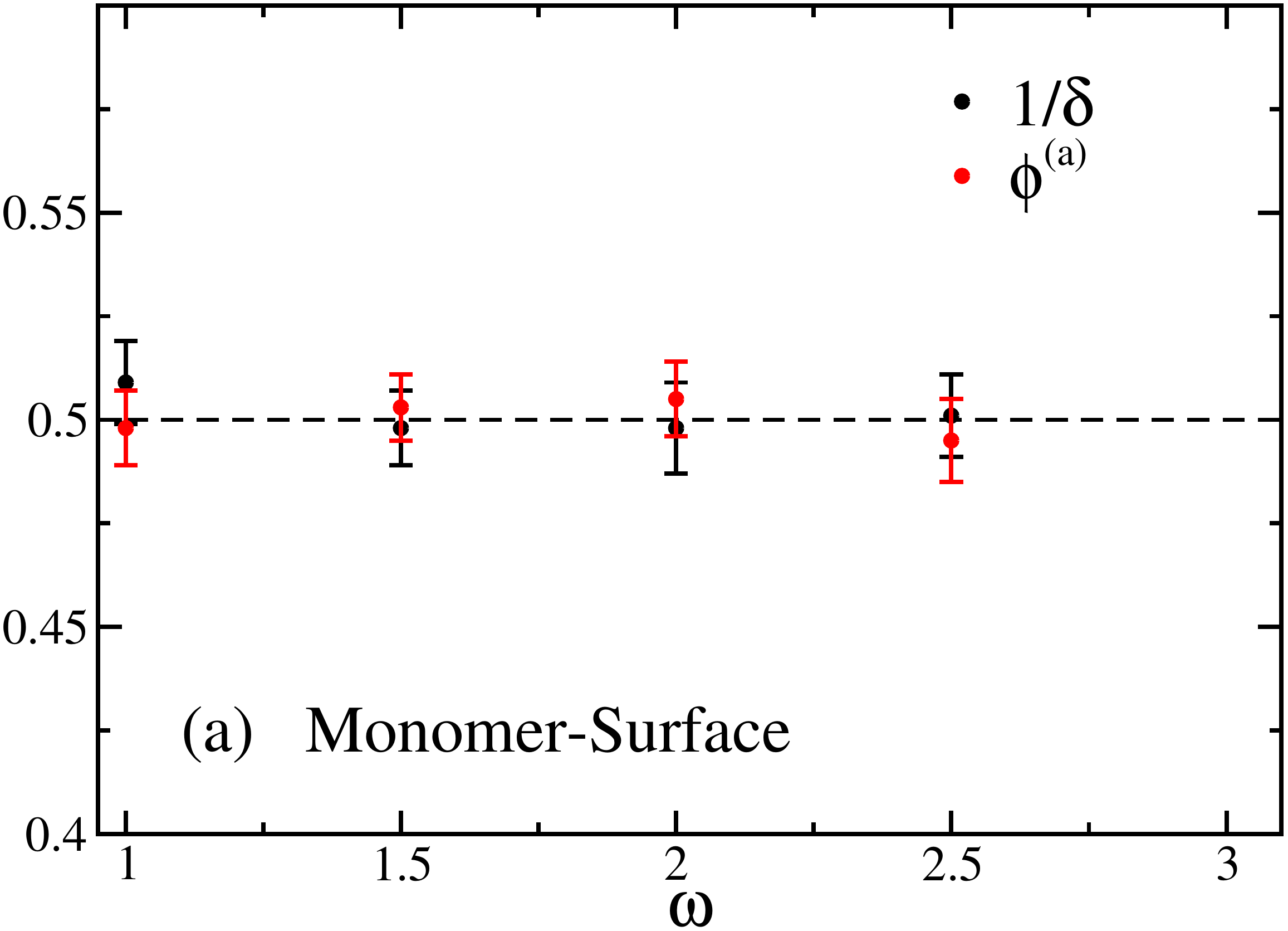}
\includegraphics[angle=0, width=0.7\columnwidth]{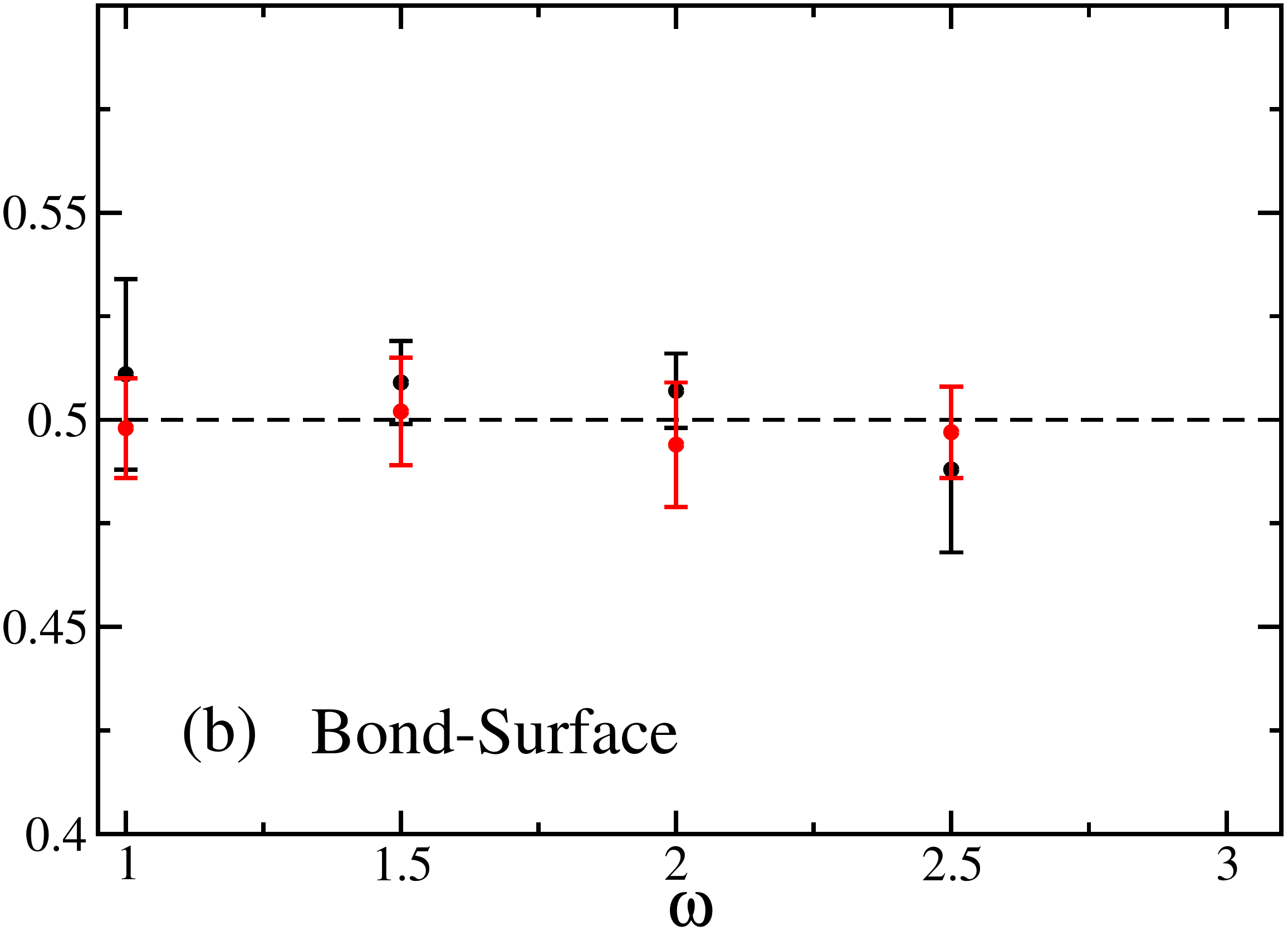}
\includegraphics[angle=0, width=0.7\columnwidth]{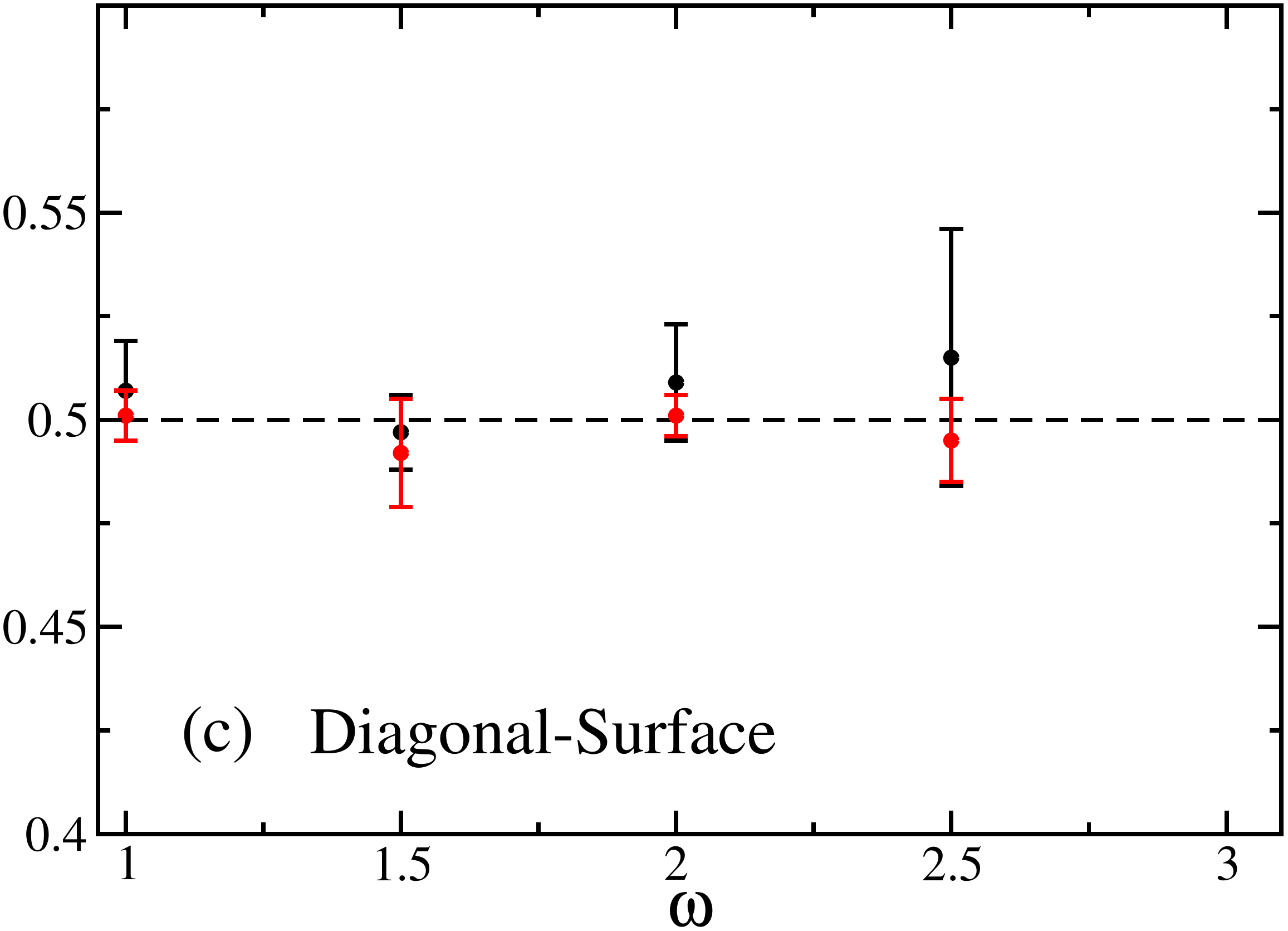}
\caption{Surface exponents as function of the bulk interaction parameter $\omega$ for trails with up to $10240$ steps. The red circles are the values for the $\phi^{(a)}$ and the black circles values for $1/\delta$. The dashed line represents the expected value in two dimensions. On (a) the MS case, (b) the BS and on (c) the DS boundary scenario.}
\label{Fig5}
\end{figure}

By performing simulations with PERM for trails with up $10240$ steps we estimate the values of $1/\delta$ and $\phi^{(a)}$ for the normal and the special surface transition for all three boundary scenarios. For the normal surface transition we investigate the four different values $\omega=1.0$, $1.5$, $2.0$, and $2.5$ in detail. In Fig.~\ref{Fig5} our estimates of the surface exponents are shown as a function of $\omega$. In all cases our estimates satisfy $1/\delta=\phi^{(a)}$ within error bars, and also agree with the expected value $\phi=1/2$ in two dimensions. We conclude that the normal surface transition shows universal behaviour as expected: the standard scaling hypothesis predicting $1/\delta=\phi^{(a)}=1/2$ is correct in the presence of attractive bulk interactions and different surface boundary conditions.

\begin{figure}[!h]
\includegraphics[angle=0, width=0.7\columnwidth]{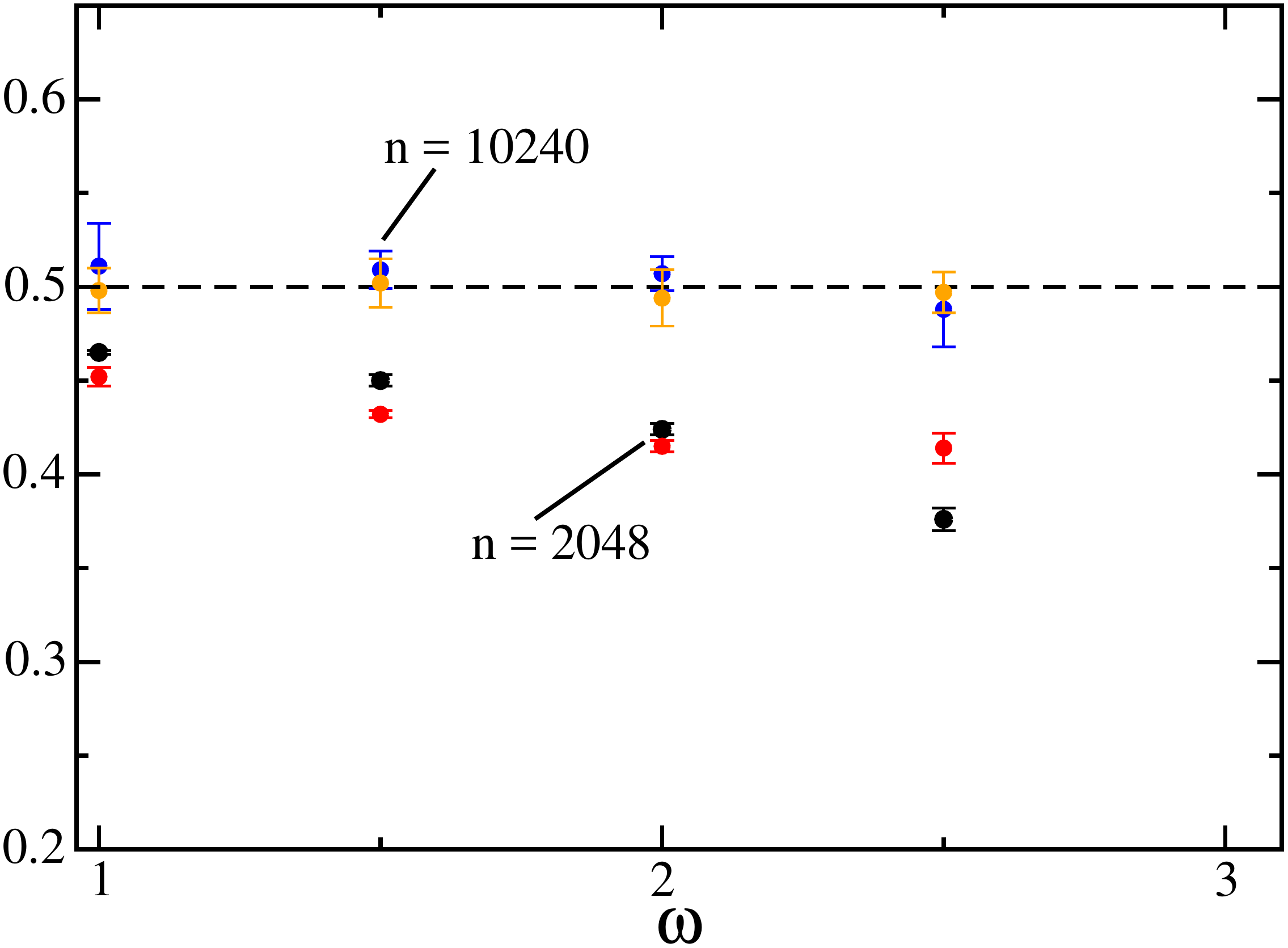}
\caption{Surface exponents $1/\delta$ and $\phi^{(a)}$ as function of $\omega$ for the BS case for trails with $2048$ and $10240$ steps. The black and red circles are the values of $1/\delta$ and $\phi^{(a)}$, respectively, for trails with $2048$ steps (lower symbols)), and the blue and orange circles are the values of $1/\delta$ and $\phi^{(a)}$, respectively, for trails with $10240$ steps (upper symbols).}
\label{Fig5x}
\end{figure}

At this point we note that we find strong corrections to scaling in the estimation of these exponents. Fig.~\ref{Fig5x} shows surface exponent estimates obtained for trails with $2048$ and $10240$ steps for the BS case. While the exponent estimates appear converged to the expected value of $1/2$ for the longer trails, there is a clear deviation for estimates from the shorter trails. Importantly, error bars at shorter lengths are misleading and would seem to support claims of non-universality.  Similar deviations are evident for the other two cases.

In the remainder of this section we discuss the special surface transition in detail. As established above, the special transition occurs at $\omega=3$ in all three scenarios, albeit at different values of $\kappa^{(s)}$. From an analysis of $R^2_{\perp/\parallel,n}$, we find $\kappa^{(s)}_{(MS)}=1.924(2)$, $\kappa^{(s)}_{(BS)}=2.442(4)$, and $\kappa^{(s)}_{(DS)}=3.001(2)$. The latter value confirms the expected exact value $\kappa^{(s)}_{(DS)}=3$; similar to the identification of $\omega^{(c)}=3$, at this value the probabilities of a stochastic growth process are perfectly balanced by the Boltzmann weight $\kappa^{(s)}_{(DS)}$. There is no known exact value for the other two cases. The BS case has been investigated previously and a value $\kappa^{(s)}_{(BS)}=2.45(5)$ was found  \cite{Foster2010}. We are not aware of any previous work regarding the value of $\kappa^{(s)}_{(MS)}$.

We note that our estimates satisfy $\kappa^{(s)}_{(MS)}<\kappa^{(s)}_{(BS)}<\kappa^{(s)}_{(DS)}$, which make sense heuristically due to the density of contacts in adsorbed configurations for each of the boundary scenarios.

\begin{figure}[!h]
\includegraphics[angle=0, width=0.7\columnwidth]{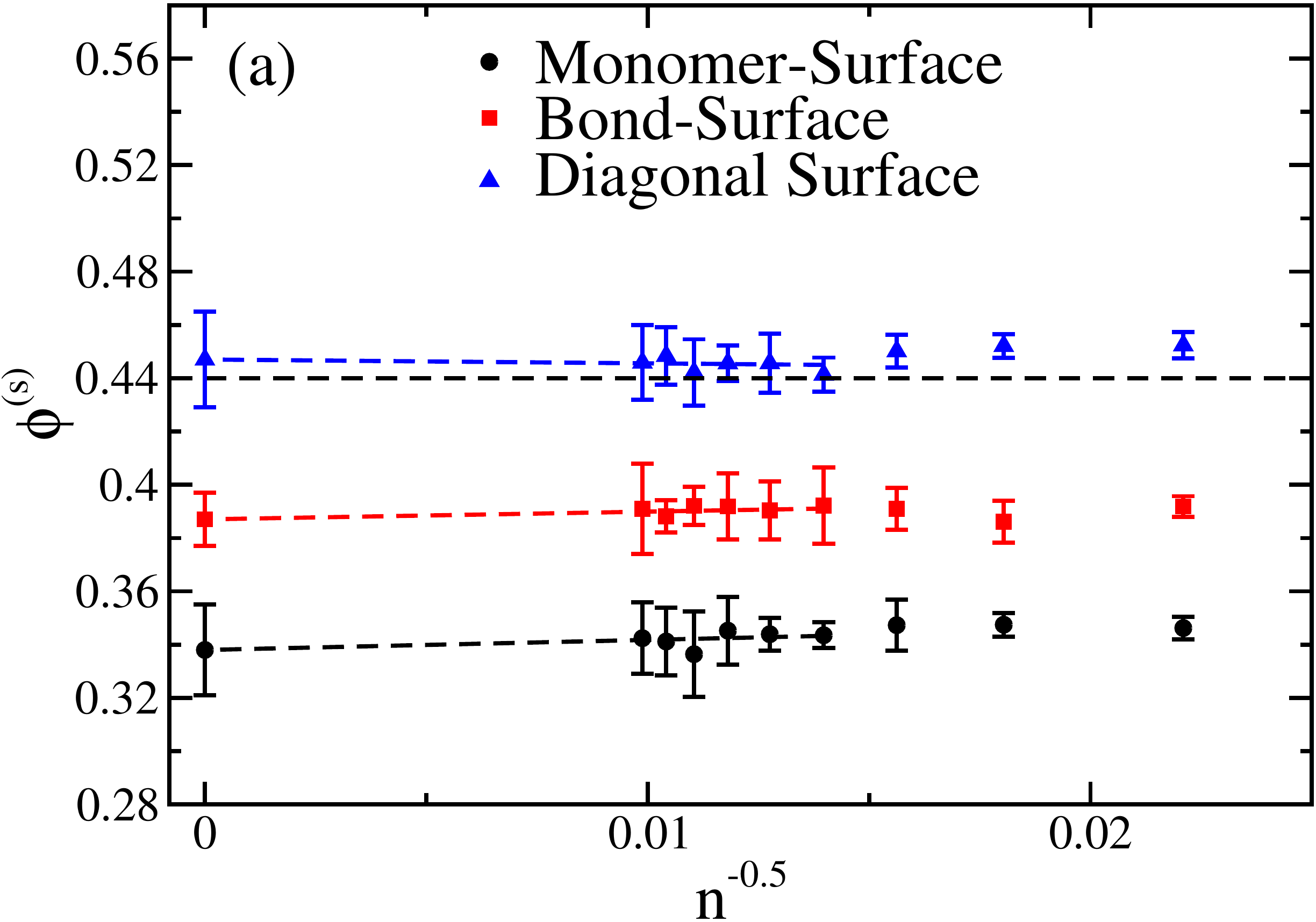}
\includegraphics[angle=0, width=0.7\columnwidth]{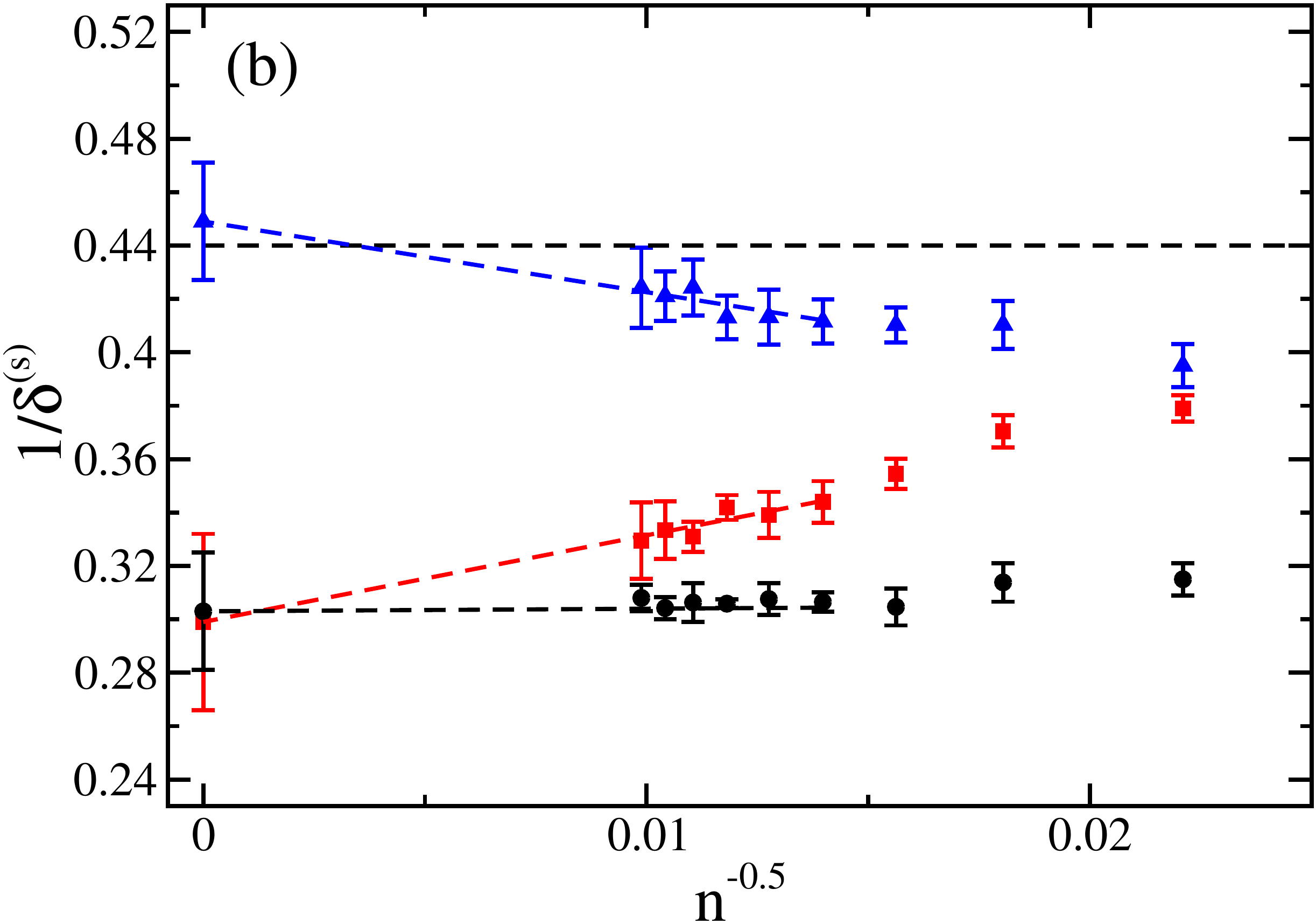}
\caption{(a) The surface exponent $\phi^{(s)}$ for different sizes as function of $n^{-0.5}$ on the special surface point. Black circles are the values for the MS case, red squares BS and blue triangles the DS case. The dashed black line is the expected value of $0.44$. (b) Exponent $1/\delta^{(s)}$ as function of $n^{-0.5}$ for the boundary scenarios.}
\label{Fig6}
\end{figure}

While we have strong confirmation of universality for the normal adsorption transition, our findings do not support universality for the special transition. Intriguingly, we find different exponent values depending on the boundary studied. Fig.~\ref{Fig6} shows finite-size estimates of $\phi^{(s)}$ (panel a) and $1/\delta^{(s)}$ (panel b) for all three boundary scenarios. The estimates for $\phi^{(s)}$ seem to have no strong size dependence, and seem to converge to three distinct values in the thermodynamic limit. We estimate $\phi^{(s)}_{(MS)}=0.338(17)$, $\phi^{(s)}_{(BS)}=0.387(10)$, and $\phi^{(s)}_{(DS)}=0.447(18)$, with the values for the DS and the BS cases being in a good agreement with those found in ~\cite{Owczarek1995,Foster2010}. The estimates for $1/\delta^{(s)}$ show a stronger size dependence. We find $1/\delta^{(s)}_{(MS)}=0.303(22)$, which is not too dissimilar from $\phi^{(s)}_{(MS)}=0.338(17)$, and $1/\delta^{(s)}_{(DS)}=0.449(22)$, which is in reasonable agreement with $\phi^{(s)}_{(DS)}=0.447(18)$. We could thus be tempted to conclude that in both of these cases the equality $1/\delta^{(s)}=\phi^{(s)}$ holds, albeit with different exponent values. However, we also find $1/\delta^{(s)}_{(BS)}=0.299(33)$, which does not support this equality. Our findings are summarised in Table \ref{tab1}.

\begin{table}[!h]
\begin{tabular}{|l||c|c|c|}\hline
&  monomer-surface & bond-surface &    diagonal surface   \\\hline\hline
$\kappa^{(s)}$  		 &   $1.924(2)$     &  $2.442(4)$  &    $3.001(2)$    \\\hline
$\phi^{(s)}$  		 &   $0.338(17)$     &  $0.387(10)$  &    $0.447(18)$    \\\hline
$1/\delta^{(s)}$     &  	 $0.303(22)$     &  $0.299(33)$  &     $0.449(22)$         \\ \hline
\end{tabular}
\caption{Values found for the surface exponents $1/\delta^{(s)}$ and $\phi^{(s)}$ for the boundary scenarios at the special surface transition point.}
\label{tab1}
\end{table}

These results do not support universality of the special surface exponents. Neither can we confirm $1/\delta^{(s)}=\phi^{(s)}$, nor seem the exponent values independent of the details of the boundary.

It is important to notice that the estimate of $\phi^{(s)}$ is highly dependent on the precise location of the adsorption point, but no such argument can be made for the method of estimating $1/\delta^{(s)}$. However, it is also important to highlight that the presence of strong finite-size corrections seriously affects exponent estimates for the ordinary surface transition, having recently led to claims of non-universality in the case of self-avoiding walks \cite{Plascak2017,Martins2018}.

\begin{figure}[!h]
\includegraphics[angle=0, width=0.7\columnwidth]{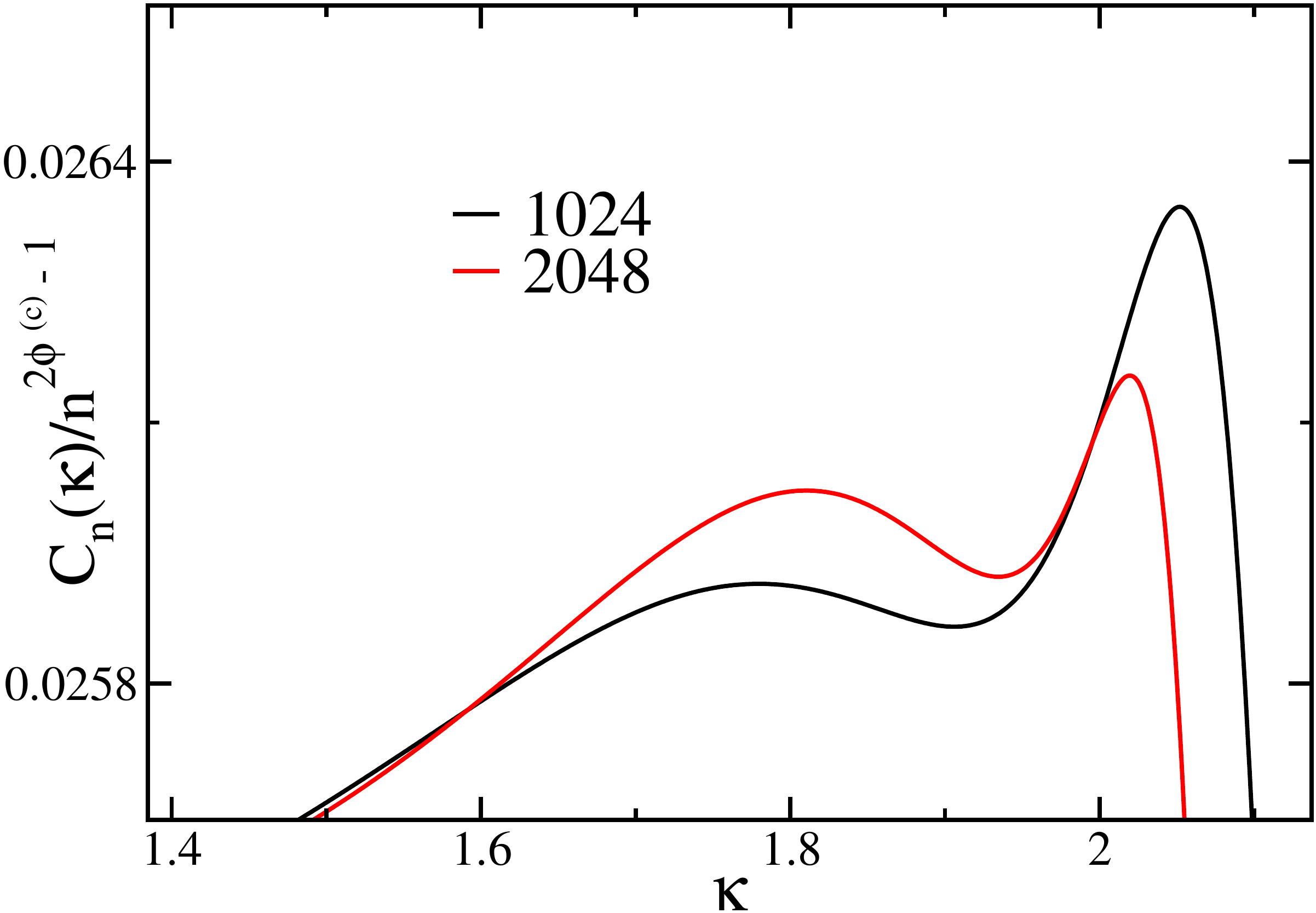}
\caption{Bulk specific heat per monomer normalized with $\phi\approx0.78$ as function of $\kappa$ for $\omega=3$ for the MS case. Two different sizes are shown, $1024$ steps (black curve) and $2048$ steps (red curve).}
\label{Fig7}
\end{figure}

We also find evidence for strong finite-size corrections at the special point. When looking more closely at the line of critical collapse approaching the special adsorption transition, we find two peaks in the bulk specific heat near the special transition, as shown in Fig.~\ref{Fig7} for the MS case. These two peaks were only found in the MS and the BS scenarios. This could be an indication of two neighbouring phase transitions. These two peaks persist in the vicinity of $\omega=\omega^{(s)}$, however one of those peaks became weaker for longer lengths ($2048$) which indicates that instead of a second phase transition we are dealing with strong finite-size corrections. While we cannot see the weakening of one of those peaks (or merging of both peaks) near the special transition, we believe that this is likely to happen for longer trails. With trails with up to $10240$ steps it was not possible to resolve this question due to statistical errors of the simulations, and we suggest that simulations with even longer configurations have to be performed to understand the precise nature of these finite size corrections and how they affect the estimates of the critical exponents.

\section{Conclusions}
\label{conclusions}

We performed simulations of the model of absorbing interacting self-avoiding trails on the square lattice using flatPERM, a uniform sampling stochastic growth algorithm. We used a two-parameter version of flatPERM to sample the density of states for trails with up $128$ steps, and a one-parameter version of flatPERM to sample trails with up to $1024$ steps, going up to $2048$ steps for specifically chosen values. We also performed PERM for trails with length up to $10240$ steps. Three different scenarios for the surface interaction were studied: monomer-surface interactions (MS), bond-surface interactions (BS) and monomer interactions at a diagonal surface (DS). 

By analysing the fluctuation map for these three scenarios we found similar phase diagrams with coil, collapsed and adsorbed phases. In all three scenarios the coil-collapsed transition was found to occur at a constant line at $\omega=3$. We also found evidence of a first-order transition between the collapsed phase and the adsorbed phase.

The main focus of this work was the analysis of adsorption from the coil phase via the normal surface transition and of adsorption from the bulk-critical phase via the special surface transition. 

We found for all three scenarios that the normal surface transition occurs along a critical line. The estimated values of the surface exponents $1/\delta$ and $\phi^{(a)}$ are both close to the expected value of $1/2$, showing that the normal surface transition is universal for trails and that the relation $\phi^{(a)}=1/\delta$ holds for different solvent conditions and different types of boundary conditions. We point out that even for trails with 2048 steps strong finite-size corrections led to exponent estimates that indicated non-universality, and that we needed to simulate consider considerably longer trails to observe the actual exponent values.

These findings are relevant with regards to the recently claimed non-universality of the adsorption transition for polymers in the presence of bulk interactions \cite{Plascak2017,Martins2018}. This was based on simulations of relatively short self-avoiding walks of lengths up to $503$ steps on the simple cubic lattice. A similar variability of exponent estimates was found for self-avoiding walks and trails in two and three dimensions for different lattices and varying interaction strengths \cite{Bradly2018,Bradly2018a}. In the latter works it was pointed out that different methods of analysis resulted in significantly different exponent estimates for configurations with steps up to length $1024$, and that choosing any single method of estimation leads to exponent estimates with erroneously small error bars. The present work indicates that increasing the size of the configurations by an order of magnitude is needed to go beyond finite-size correction terms which are seemingly not captured by any of the methods used in estimating the exponents.

When analysing the special adsorption transition for $\omega=3$, we found critical behaviour in all three scenarios. 
One of the aims of the present study was to investigate the discrepancy between the previously reported values of $\phi^{(s)}$ for the (DS) and (BS) cases.
Our estimates, summarised in Table \ref{tab1}, do not resolve 
this discrepancy. In addition, for the (MS) case we find yet another value of $\phi^{(s)}$.
Moreover, our estimates of $1/\delta^{(s)}$ deviate from the respective values of $\phi^{(s)}$ with the exception of the (DS) case, where we find good agreement within error bars.

We would like to argue that the discrepancy between all of these estimates is likely due to very strong finite-size corrections to scaling around a higher order critical point. Support for this comes from the observation that near this point we find two weak but clearly separated peaks in the bulk-specific heat, which seem to weaken and move closer to each other at longer lengths. This behaviour is not captured by our scaling assumptions, and hence one needs to either amend these assumptions to capture this behaviour, or to perform simulations at even longer lengths than $10240$ steps to get beyond these corrections to scaling.

As the simulations in ~\cite{Owczarek1995} have been performed for $10^6$ steps at the exact value $\kappa^{(s)}=3$, and as this is the only scenario for which in the present work we find agreement between $\phi^{(s)}$ and $1/\delta^{(s)}$, we assert that most likely the special surface transition is universal and that the associated surface exponent equals $\phi^{(s)}=1/\delta^{(s)}=0.45(2)$. If there existed different sets of surface exponents for the special transition, then one would have to still identify a mechanism that would be capable of inducing this difference. We cannot discount this completely, as collapsing trails likely have a length scale exponent $\nu=1/2(\log)$, and in a scaling limit the lattice structure near the boundary only becomes irrelevant if $\nu>1/2$. One possible mechanism to cause a change, suggested to us by Tiago Jos\'e de Oliveira, is that for the horizontal boundary long adsorbed segments of the trail in the surface suppress interactions in the layer above, which is not the case for the DS case. More work is needed to resolve this issue.

\acknowledgments

N. T. R. thanks Queen Mary University London for hosting while this work was carried out, and gratefully acknowledges use of the university's HPC cluster. N. T. R also acknowledges financial support from CAPES. Financial support from the Australian Research Council via its Discovery Projects scheme (DP160103562) is acknowledged by A. L. O. The authors are grateful to Tiago Jos\'e de Oliveira for helpful comments and critical reading of the manuscript.

\end{document}